\documentclass{aastex62}
\usepackage{bm}
\usepackage{amsmath}
\usepackage{listings}
\definecolor{dkgreen}{rgb}{0,0.6,0}
\definecolor{gray}{rgb}{0.5,0.5,0.5}
\definecolor{mauve}{rgb}{0.58,0,0.82}
\graphicspath{{./}{figures/}}

\received{---}
\revised{---}
\accepted{---}
\submitjournal{ApJs}

\begin{document}

\title{Resistive and multi-fluid RMHD on graphics processing units}

\correspondingauthor{A. J. Wright}
\email{A.J.Wright@Soton.ac.uk}

\author[0000-0002-5953-4221]{A. J. Wright}
\affiliation{Mathematical Sciences and STAG Research Centre,\\
	University of Southampton SO17 1BJ, United Kingdom}
\affil{Next Generation Computational Modelling Group, \\
University of Southampton SO16 7PP, United Kingdom}
\nocollaboration

\author[0000-0003-4805-0309]{I. Hawke}
\affiliation{Mathematical Sciences and STAG Research Centre,\\
University of Southampton SO17 1BJ, United Kingdom}
\affil{Next Generation Computational Modelling Group, \\
	University of Southampton SO16 7PP, United Kingdom}
\nocollaboration



\begin{abstract}

In this work we present a proof of concept of CUDA-capable, resistive, multi-fluid models of relativistic magnetohydrodynamics (RMHD). Resistive and multi-fluid codes for simulating models of RMHD suffer from stiff source terms, so it is common to implement a set of semi-implicit time integrators to maintain numerical stability. We show, for the first time, that finite volume IMEX schemes for resistive and two-fluid models of RMHD can be accelerated by execution on graphics processing units, significantly reducing the demand set by these kinds of problems. We report parallel speed-ups of over 21$\times$ using double precision floating-point accuracy, and highlight the optimisation strategies required for these schemes, and how they differ from ideal RMHD models. The impact of these results is discussed in the context of the next-generation simulations of neutron star mergers.

\end{abstract}



\section{Introduction}\label{sec:intro}
Graphics processing units (GPUs) have been steadily gaining attention in the scientific community for a number of years. As the availability of large numbers of GPUs in high performance computing clusters is growing, and the peak theoretical performance of graphics cards is increasing exponentially, GPUs are being adopted to solve compute-intensive tasks.

GPUs were initially developed for graphics rendering---the simple, independent instructions required for this process are data-parallel, and additional hardware was designed to meet these needs. What graphics cards sacrifice in single thread execution, they make up for in sheer numbers of compute cores, allowing massively parallel tasks to be completed rapidly. The introduction of the compute unified device architecture (CUDA) \citep{Nickolls2008} developed by NVIDIA made accessible their GPUs for general purpose programming, and have since seen use in a wide range of scientific applications from neuroimaging \citep{Lee2012} to deep learning \citep{Gawehn2018}. In this work, we are interested in the applicability of GPUs in evolving different models of  relativistic magnetohydrodynamics (RMHD). 

Many astrophysical simulations require the evolution of magnetofluids, and by and large these systems are most commonly described by ideal MHD \citep{Kiuchi2015a, Ruiz2018, Nouri2018}. Ideal MHD describes a single-fluid, perfectly conducting, charged plasma, and the resulting equations of motion take a simple, balance law form \citep{Anton2010}. As a result, the standard methods for evolving systems described by ideal MHD are very efficient, and the amount of computation required per timestep is relatively low compared to alternative models. However, whilst there is still a huge amount of success to be had in modelling events such as neutron star mergers using this description \citep{Font2008, Koppel2018, Radice2017a}, essential physics may be being lost. The work of \citet{Dionysopoulou2015, Dionysopoulou2013} suggests that the dynamics of mergers can be altered significantly by generalising current models to include finite conductivities. These resistive models of MHD can be taken one step further, as suggested in \citet{Andersson2016, Andersson2016a, Andersson2016b}, and extended to describe multiple interacting charged fluid species, coupled through electromagnetism. 

Multi-fluid models of resistive MHD provide a more accurate description of these kinds of systems. Multi-fluid effects such as entrainment and the two-stream instability cannot be captured using the single fluid, ideal approximation, and instabilities have been shown to have a major impact in the evolution of neutron star mergers \citep{Price2006, Obergaulinger2010, Kiuchi2015}. Of course, there are difficulties that lay ahead in modelling mergers using these methods. Multi-fluid models inherently contain larger state vectors than single fluid models, and therefore require additional computation per timestep. Furthermore, the multi-fluid description proposed in \cite{Andersson2016b} is not currently in a suitable form for numerical evolution. However, we expect the form of the equations to resemble a balance-law form with the some, potentially very stiff, source term. To prevent numerical instabilities arising from stiff source contributions, it is common to employ implicit time-integrators of the type in \cite{Pareschi2004}. These kinds of integrators, whilst they reduce the total execution time that would be achieved with explicit integrators in many regimes, can make these models too slow for practical application. 

If resistive, multi-fluid descriptions of MHD for merger simulations are to be achieved, methods need to be developed that will significantly reduce the computational demands set by these models. Investigations into GPU-implementations of ideal MHD models shows promise, and significant performance improvements are reported in \cite{Zink2011} and \cite{Wong2011}. Until now, there has been no investigation into the execution of alternative models of MHD on graphics cards.  

A multidimensional, general-relativistic formulation of four-fluid MHD, as proposed in \cite{Andersson2016}, would require the evolution of over 30 variables, and a huge amount of computation at each timestep when compared to ideal single fluid models---which themselves require the evolution of less than 10 variables. Fortunately, however, the system of equations is highly parallelisable. The time integration for each computational cell is independent, and the flux methods depend upon only a handful of neighbours, meaning that there is a huge potential for speed-ups using a large number of compute cores. 

A major barrier to improving speed-ups using massively parallel processors, however, lies in their memory limitations. Whilst there is the potential to execute a large amount of independent computation in parallel over many thousands of cores, whether the substantially increased size of the multi-fluid system will affect the available speed-up, and whether the limitations on the size of the fastest memory on the GPU will become significant, is unclear. It is the purpose of this work to assess whether a possible avenue to efficient multi-fluid models of RMHD is to utilise graphics processing units to significantly reduce computation times. To do this we present METHOD, a new open-source (see section \ref{sec:discussion}) multi-fluid, electro-magnetohydrodynamics code that we use to demonstrate significant performance improvements available for resistive, single and two-fluid models of RMHD by utilising GPUs. 

The rest of the this paper is laid out as follows: In section \ref{sec:models and methods} we describe the models that are implemented in METHOD, and the numerical methods we use to solve the equations of motion; section \ref{sec:METHOD} regards how METHOD is designed and the optimisation strategies implemented to maximise performance; in section \ref{sec:validation and performance} we validate METHOD via comparisons with exact solutions and convergence tests, and report the parallel speed-up on different GPUs; finally, in section \ref{sec:discussion}, we end with a discussion of how this work relates to the future of neutron star merger simulations.

\section{Models and numerical methods}\label{sec:models and methods}

METHOD is a multi-physics, numerical code, implementing three approximations of multi-fluid RMHD. Numerically, plasma dynamics differ from fluid dynamics in the restrictions set by Maxwell's equations, specifically the divergence constraints of Gauss for the electric and magnetic fields. To satisfy these constraints we use the method of hyperbolic divergence cleaning for each model \citep{Komissarov2008}. It is worth pointing out that \cite{Amano2016} uses the constrained transport method for the two-fluid RMHD model, making the argument that divergence cleaning may not be suitable. We have found that, for the applications in which we use METHOD, divergence cleaning works well, and that implementing this method requires very little change to any existing code.

As we believe that multi-fluid descriptions of MHD will be required for only the smallest scales of a neutron star merger simulation, for the purpose of this work we limit ourselves to local regions of spacetime, and thus to special relativistic descriptions. Therefore, we assume a flat Minkowski geometry (i.e.\ special relativity), use natural units where the speed of light $c=1$ and absorb the permittivity and permeability, $\epsilon_0$ and $\mu_0$, into the definition of the electromagnetic fields. We also adopt the Einstein summation convention over repeated indices unless explicitly stated otherwise, and use Latin indices for three spatial dimensions.

\subsection{Plasma models}
There are three distinct plasma models implemented in METHOD: ideal RMHD; resistive, single fluid RMHD; and resistive, two-fluid RMHD. As this paper reports on the benefits of GPU implementations for resistive RMHD, we will only briefly describe the resistive models, as the details can found in more detail in the references. For each model, we assume special relativity, use divergence cleaning for Maxwell's constraints, and perform the flux calculations using flux vector splitting and an asymptotically third-order WENO3 reconstruction (section \ref{sec:FVS}). 

Each model requires an equation of state to close the system of equations to relate the hydrodynamics quantities. As is common, we adopt the $\Gamma$-law equation of state, $p=\rho e (\Gamma - 1)$.

\subsubsection{Single fluid, resistive RMHD}
The formulation of the resistive model can be found in more detail in \cite{Komissarov2008} and \cite{Dionysopoulou2013}, along with a motivation for the choice of Ohm's law which we use throughout this work. The system of conservation equations has the following form:
\begin{eqnarray} \label{eq:resistive RMHD}
\partial_t  \left(
\begin{array}{c}
D \\ S_i \\ \tau \\ B^i \\ E^i \\ \varrho \\ \psi \\ \phi
\end{array} \right)
+ \partial_k  \left(
\begin{array}{c}
Dv^k \\ S^k_i \\ S^k - D v^k \\ -\epsilon^{ijk} E_j + \delta_i^k \phi\\ \epsilon^{ijk} B_j + \delta_i^k \psi \\ J^k \\ E^k \\ B^k
\end{array} \right)
=  \left(
\begin{array}{c}
0 \\ 0 \\ 0 \\ 0\\ -J^i \\ 0 \\ \varrho - \kappa \psi \\ -\kappa \phi
\end{array} \right)
\end{eqnarray}
where we relate the conserved vector, $\bm{q}=(D, S^i, \tau, B^i, E^i, \varrho)$, to the primitive quantities, $\bm{w}=(\rho, v^i, p, B^i, E^i)$, using the following,
\begin{eqnarray} \label{eq:resistive prims}
 \left(
\begin{array}{c}
D \\ S_i \\ \tau \\ J_i \\ S_{ij} \\ 
\end{array} \right)
=  \left(
\begin{array}{c}
\rho W \\ \rho h W^2 v_i + \epsilon_{ijk}E^j B^k \\ \rho h W^2 - p + \frac{1}{2}(E^2 + B^2) - \rho W \\ \varrho v_i + W \sigma [E_i + \epsilon_{ijk}v^j B^k - (v_k E^k)v_i] \\ \rho h W^2 v_i v_j + [p + \frac{1}{2}(E^2 + B^2)] \delta_{ij} - E_i E_j - B_i B_j
\end{array}  \right),
\end{eqnarray}
and define the Lorentz factor and specific enthalpy as
\begin{eqnarray}
W &= 1/\sqrt{1 - v_i v^i} \\
h &= 1 + \frac{\Gamma p }{ \rho (\Gamma-1)}.
\end{eqnarray}
Throughout this paper, we use $\{\rho, v_i, p, e, E_i, B_i\}$ to represent the rest-mass density (measured in the rest-frame of the fluid), velocity (in the lab frame) in the $i^{\text{th}}$-direction, hydrodynamic pressure, specific internal energy, and the electric and magnetic fields in $i^{\text{th}}$-direction respectively.

The conserved quantities $\{D, S^i, \tau\}$ represent the relativistic mass-energy density, momentum density in the i$^{\text{th}}$ direction, and a kinetic energy density-like term, and we make the choice to evolve the charge density, $\varrho$ (initialising it through Gauss' law, $\nabla \cdot E = \varrho$) so as to avoid error accumulation from multiple finite-differencing in the case of discontinuous electric fields. We adopt the same definition of the charge density current as \cite{Dionysopoulou2013} and \cite{Palenzuela2009}, with $J^i = \varrho v^i + W \sigma [E^i + \epsilon^{ijk} v^j B^k - (v_k E^k)v^i]$.

Finally, the electric conductivity is given by $\sigma$, which in the ideal RMHD limit, $\sigma \to \infty$. This leads to numerical difficulties for systems with high conductivities, as the source terms for the electric fields become large. To maintain stability and reduce the restriction set by the CFL condition, it is common to utilise semi-implicit integration schemes such as those of section \ref{sec:IMEX} for stiff systems \citep{Dionysopoulou2013, Palenzuela2009, Dionysopoulou2015}.

The final two equations in (\ref{eq:resistive RMHD}) are a consequence of the divergence cleaning method \citep{Dedner2002}. In METHOD, we set the timescale $\kappa^{-1}=1$ for all simulations, and initialise the scalar fields $\phi(t=0)=0$ and $\psi(t=0)=0$.

\subsubsection{Two-fluid, resistive RMHD}
The resistive two-fluid description closely follows that presented in \cite{Amano2016} and \cite{Balsara2016}. We describe two charged species, denoted by $s\in[p, e]$ for protons/positrons and electrons respectively, with charge-mass ratios given by $\mu_s = q_s/m_s$ (where there is no sum over species, $s$). 

The equations of motion of the two-fluid system describe how the total, hydrodynamic quantities evolve, and how the same quantities weighted with respect to the charge-mass ratio evolve---the sum weighted with respect to the charge-mass ratio is chosen as it naturally gives rise to charge conservation and a generalised Ohm's law, as we shall see.  It is therefore convenient to define an implied sum over species, in which we assume the total contribution of some variable, $U$, due to the two charge species is given as $U = \sum_s U_s = U_e + U_p := U_s$. For clarity, the conserved quantity, $D$, is therefore given by the sum of the contributions, $D = D_e + D_p = \sum_{s} D_s := D_s = \rho_s W_s$.

In the same form as the single fluid equations, we get the standard balance-law form,
\begin{equation}\label{eq:cons law}
\partial_t \bm{q} + \partial_j \bm{f}^j(\bm{q}) = \bm{s}(\bm{q}),
\end{equation}
in which the conserved vector, $\bm{q}=\{D, S_i, \tau, \overline{D}, \overline{S}_i, \overline{\tau}, B^i, E^i, \psi, \phi\}$ is related to the primitive quantities through
\begin{eqnarray} \label{eq:twoFluid cons}
\left(
\begin{array}{c}
D \\ S_i \\ \tau \\ \overline{D} \\ \overline{S}_i \\ \overline{\tau}
\end{array} \right) = \left(
\begin{array}{c}
\rho_s W_s \\ \rho_s h_s W_s^2 v_{s,i} + \epsilon_{ijk} E^j B^k \\ (\rho_s h_s W_s^2 - p_s) + (E^2 + B^2) / 2 - \rho_s W_s \\ \mu_s \rho_s W_s \\ \mu_s \rho_s h_s W_s^2 v_{s,i} \\ \mu_s (\rho_s h_s W_s^2 - p_s - \rho_s W_s)
\end{array} \right),
\end{eqnarray}
where we are assuming the sum over fluid species, $s$, and the primitive quantities have the same interpretation as in the single fluid case, but specific to a charged species. We present the flux and source vectors in appendix \ref{sec:two fluid features}.

As in the single fluid case, the conserved quantities $\{D, S^i, \tau\}$ represent the relativistic mass-energy density, momentum density, and kinetic energy-like density---i.e.\ the sum of contributions of each fluid species. The barred alternatives, $\{\overline{D}, \overline{S}^i, \overline{\tau}\}$, correspond to the weighted sum with respect to the charge-mass ratio, excluding any contribution from the electromagnetic fields . 

\begin{figure}[!h]
	\plottwo{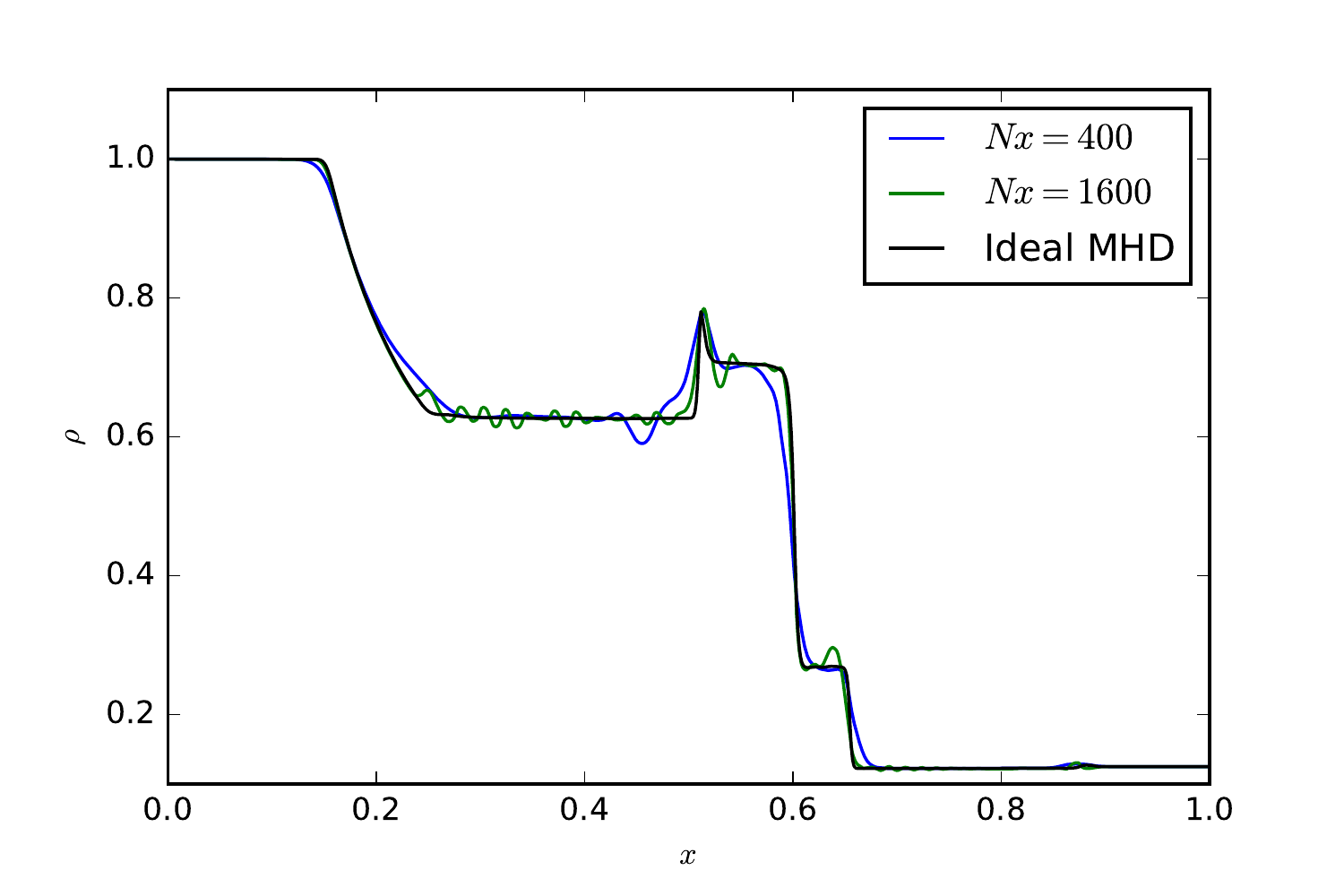}{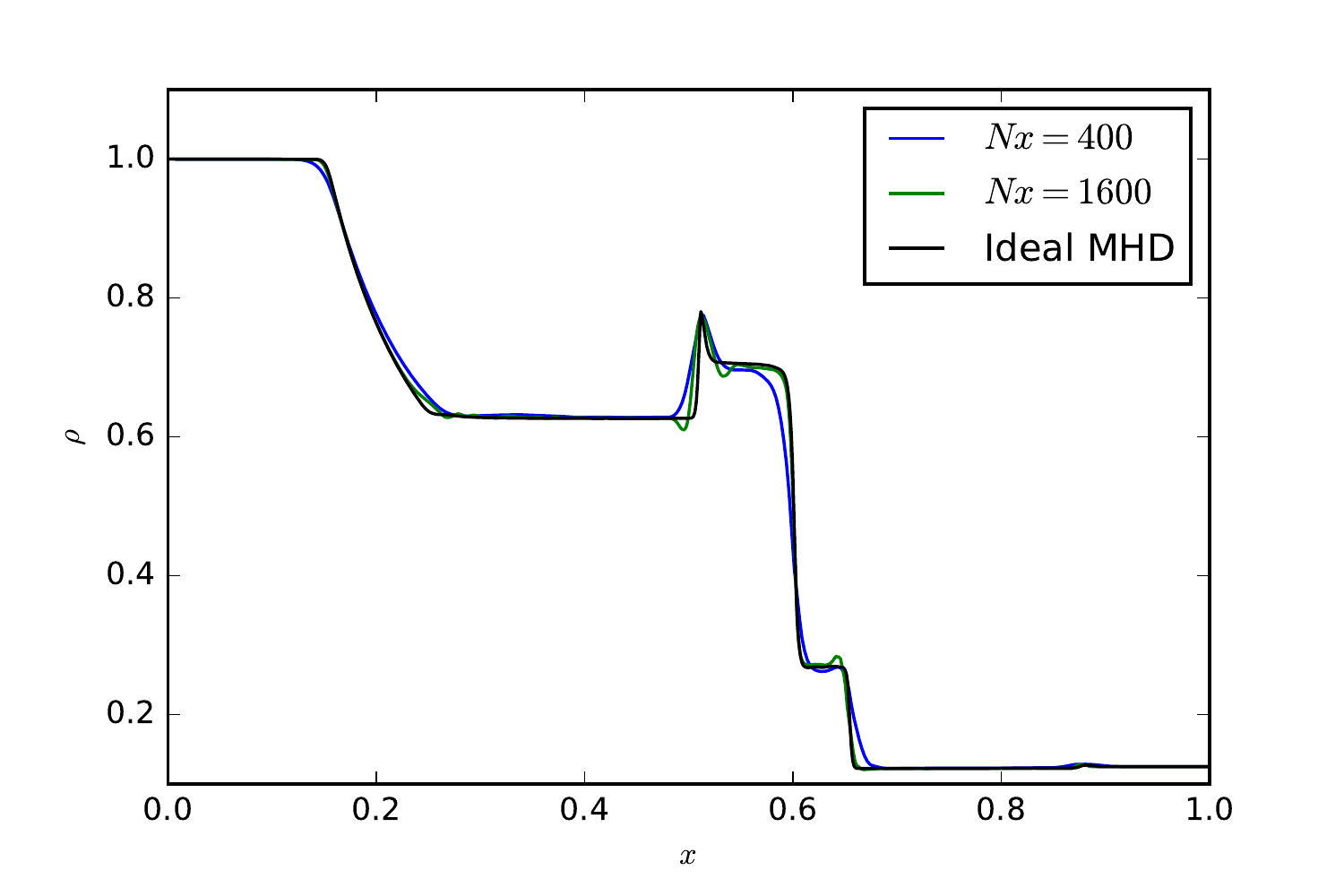}
	\caption{The rest-mass density for the two-fluid Brio-Wu shock tube test. Left: the results for the case of an electron-positron, pair plasma, i.e. $\mu_p = 10^3 = −\mu_e$. Right: an electron-proton plasma of $\mu_p = 10^3 = \mu_e/100$. For both figures, the solution generated by the single fluid, ideal MHD model is shown for reference. We have used up to 1600 grid points and evolved the system until $T = 0.4$. Due to the high resolution, we can see the two-fluid effect manifest as perturbations upon the single fluid result---this effect is more pronounced for the higher resolution run as the effect is more easily resolved, and less pronounced for the non-pair plasma (right) due to the reduced skin depth (by a factor of 100$\times$).}
	\label{fig:brio wu two fluid}
\end{figure}

A note on the plasma frequency, $\omega_p = \sqrt{\mu_s^2 \rho_s} $. A key factor differentiating single and multi-fluid plasmas is one of charge separation, and the so-called `two-fluid effect' which manifests as oscillations of a fluid's constituent charged species. The degree to which this effect can be seen is dependant upon the spatial resolution of the simulation and the plasma skin depth, $\lambda_p = 1 / \omega_p$. That is, we expect to see differences between single and multi-fluid descriptions when the spatial resolution is smaller than the skin depth, or $\Delta x < \lambda_p$, as will be seen in figure \ref{fig:brio wu two fluid}.

\subsubsection{Primitive recovery} \label{sec:primitive recovery}
An important way in which Newtonian and relativistic MHD differ is in the transformation from the evolved, conserved variables, $\bm{q}$, and the primitive variables, $\bm{w}$. In the relativistic regime, there is no closed form for the primitive quantities in terms of the conserved, rather, one must perform a root-finding procedure to determine the values of $\bm{w}$ that give rise to the current state of the system, $\bm{q}$. 

There are a number methods for this, see e.g.\ \citet{Amano2016, Dionysopoulou2013} and \citet{Palenzuela2009}, of which we have implemented the latter two. In contrast to ideal RMHD, which requires a two-dimensional root-find to determine the primitive variables \citep{Anton2010}, resistive RMHD only requires a single residual to be minimised. In order to minimise the residuals given in \cite{Dionysopoulou2013} and \cite{Palenzuela2009}, we use Newton-Raphson gradient decent. We find that in some instances one method may converge whereas the other may not, and so implementing both residuals allows us to evolve a system with more confidence that the primitive variables will be found. However, in the vast majority of cases either method will suffice. 

Whichever primitive recovery method is implemented, the absolute tolerance to which the residual is to be minimised is important. For example, when evolving the two-fluid model, any errors due to a tolerance of no lower than $10^{-12}$ propagate quickly, producing unphysical solutions and ultimately killing the simulation. It is for this reason that the tolerance is set to $10^{-13}$ for all primitive recovery procedures in METHOD. This limit on the tolerance has implications on the possible accelerations we can achieve through optimisations, see section \ref{sec:optimisations}.

In order to proceed as in the resistive, single fluid RMHD case for the two-fluid model, we must first decompose the contribution to the conserved fields into contributions from each species. For the relativistic mass-energy density, $D$, we see that
\begin{equation}
D_e = \frac{ \bar{D} - \mu_p D}{\mu_e - \mu_p}, \text{ \ \ and  \ \ }
D_p = \frac{ \bar{D} - \mu_e D}{\mu_p - \mu_e}.
\end{equation}
For the momentum and kinetic energy terms, we must first subtract the EM field contribution as is normal in resistive, single fluid RMHD. For example, we can define $\tilde{S}_i = \bar{S}_i - \epsilon_{ijk} E^j B^k$ and then proceed as before with
\begin{equation}
\tilde{S}_{i,e} = \frac{ \bar{S_i} - \mu_p \tilde{S}_{i}}{\mu_e - \mu_p}, \text{ \ \ and  \ \ }
\tilde{S}_{i,p} = \frac{ \bar{S_i} - \mu_e \tilde{S}_{i}}{\mu_p - \mu_e}.
\end{equation}
Now, $\tilde{S}_{s,i}$ represents the purely hydrodynamical momentum of species $s$ in the $i^{\text{th}}$ direction, i.e.\ $\tilde{S}_{s,i}=\rho_s h_s W_s^2 v_{s,i}$.

With these, we can continue the primitive recovery as in the resistive, single fluid RMHD for each fluid species separately. We shall see that this algorithm demands a significant amount of computation, especially when incorporated with the IMEX schemes of the next section.

\subsection{System evolution}
Here, we detail the numerical schemes that solve the equations of motion presented in the previous sections. METHOD uses a high resolution shock-capturing approach, and provides a number of different options for how to evolve a simulation. In what follows, we will only describe those methods used in generating the results for this work.
\subsubsection{IMEX schemes}\label{sec:IMEX}
The equations of motion for both resistive models resemble a conservation equation with the addition of a (possibly) stiff source term---$\sigma \to \infty$ for the single fluid description, and $\sigma \to 0$ for the two-fluid. This is also the case for more general multi-fluid models that are of interest to neutron star simulations \citep{Andersson2016, Andersson2016a, Andersson2016b}. In general, these kinds of systems will require the use of implicit time integrators to allow for the management of the large source terms. We implement the implicit-explicit, strong stability preserving, Runge-Kutta integrators of \cite{Pareschi2004}. 

Redefining the systems (\ref{eq:resistive RMHD}) and (\ref{eq:cons law}) as 
\begin{equation} \label{eq: relaxation form}
\partial_t \bm{q} = \bm{\mathcal{L}}(\bm{q}) + \bm{\Psi}(\bm{q}),
\end{equation}
where we have defined the numerical flux function, $\bm{\mathcal{L}}(\bm{q}) = -\partial_k \bm{f}^k(\bm{q})$, and the source term as $\bm{\Psi}(\bm{q})$, an IMEX scheme takes the general form, 
\begin{eqnarray}
\bm{q}^{(i)} = \bm{q}^n + \Delta t \sum_{j=1}^{i-1} \tilde{a}_{ij} \bm{\mathcal{L}}(\bm{q}^{(i)}) + \Delta t \sum_{j=1}^{\nu} a_{ij} \bm{\Psi}(\bm{q}^{(i)}), \label{eq:implicit stage} \\
\bm{q}^{n+1} = \bm{q}^n + \Delta t \sum_{j=1}^{\nu} \tilde{w}_j \bm{\mathcal{L}}(\bm{q}^{(i)}) + \Delta t \sum_{j=1}^{\nu} w_{j} \bm{\Psi}(\bm{q}^{(i)}). \label{eq:correction stage}
\end{eqnarray}

The matrices $\tilde{A}$ and $A$, with components $\tilde{a}_{ij}$ and $a_{ij}$, are constructed in such a way that the resulting equations for the flux contribution are explicit, and conversely are implicit for the source terms. Implicit and explicit forms for Equation (\ref{eq:implicit stage}) are enforced by ensuring non-zero and zero diagonal components in the respective matrices, $A$ and $\tilde{A}$---therefore, $a_{ij} = 0$ for $i>j$. and $\tilde{a}_{ij} = 0$ for $i\ge j$

The coefficients are derived by comparing the numerical solution with the Taylor expansion of the exact solution, as in other RK methods. Using the standard Butcher notation (Table \ref{table:doubleTableu}) 
\begin{table}[h!]
	\centering
	\caption{General Butcher notation for Runge-Kutta schemes.}
	\label{table:doubleTableu}
	\begin{tabular} { l|p{1cm} }
		$\tilde{c}$ &\ \ $\tilde{A}$ \\
		\hline 
		& \ $\tilde{w}^T$
	\end{tabular}
	\qquad
	\begin{tabular} { l|p{1cm} }
		$c$ &\ \ $A$ \\
		\hline 
		& \ $w^T$
	\end{tabular}
\end{table} 
the second order accurate SSP2(222), and third order accurate SSP3(332) coefficients are given Table \ref{table:SSP2222coefs} and Table \ref{table:SSP3332}, respectively, where $\gamma = 1 - 1/\sqrt{2}$, and the notation SSP$k(s, \sigma, p)$ refers to the order of the explicit part, $k$, number of implicit and explicit stages, $s$ and $\sigma$, and order of the IMEX scheme, $p$. 

\begin{table} [h!]
	\centering
	\caption{The Butcher notation for second-order SSP-RK coefficients.}
	\label{table:SSP2222coefs}
	\begin{tabular} { l|p{1cm} p{1cm} }
		$0$ & $0$ &$0$ \\
		$1$ & $1$ &$0$ \\
		\hline 
		& $1/2$& $1/2$ 
	\end{tabular}
	\qquad
	\begin{tabular} { c|cc }
		$\gamma$ & $\gamma$& $0$\\
		$1-\gamma$ & $1-2\gamma$ &$\gamma$ \\
		\hline
		& $1/2$ &$1/2$
	\end{tabular}
\end{table}
\begin{table} [h!]
	\centering
	\caption{The Butcher notation for third-order SSP-RK coefficients.}
	\label{table:SSP3332}
	\begin{tabular} { l|p{1cm} p{1cm} p{1cm} }
		$0$ & $0$ &$0$ &$0$ \\
		$1$ & $1$ &$0$ &$0$ \\
		$1/2$ & $1/4$ & $1/4$ & $0$ \\
		\hline 
		& $1/6$& $1/6$ &$2/3$
	\end{tabular}
	\qquad
	\begin{tabular} { c|ccc}
		$\gamma$ & $\gamma$& $0$ &$0$\\
		$1-\gamma$ & $1-2\gamma$ &$\gamma$ &$0$\\
		$1/2$ & $1/2 - \gamma$& $0$ &$\gamma$ \\
		\hline
		& $1/6$ &$1/6$ &$2/3$
	\end{tabular}
\end{table}

The implicit stages given by Equation (\ref{eq:implicit stage}) require a root-finding procedure of the same dimensionality as the conserved vector $\bm{q}$. A guess is made for the solution $\bm{q}^{(i)}$, say $\bm{q}^{(*)}$, and the difference of the left- and right-hand sides of Equation (\ref{eq:implicit stage}) is computed as the residual to minimise. We minimise Equations (\ref{eq:implicit stage}) using the CMINPACK library, which we discuss in more depth in section \ref{sec:optimisations}.

To determine the flux and source vectors for this guess at each iteration we require the values of the primitive quantities that correspond to the state $\bm{q}^{(*)}$. This process itself requires a root-finding procedure as mentioned in section \ref{sec:primitive recovery}. As a result, the majority of the simulation time is spent performing the conservative to primitive variable transformation. This is important when optimising the performance of the code.

\subsubsection{Flux reconstruction} \label{sec:FVS}
There are numerous procedures for determining the numerical flux function, $\bm{\mathcal{L}}(\bm{q})$, with varying complexities and advantages/disadvantages. Common methods, such as the HLL-type schemes, are widely used but can become computationally expensive when moving to higher orders. Therefore, to reduce the cost of the reconstruction and to allow an easy extension to multiple dimensions and multiple physics models, we implement the method of flux vector splitting (FVS), proposed in \cite{Shu1997}.

In FVS, the flux of the conserved quantities through cells is considered to be composed of purely left- and right-going components. The components are then interpolated across a section of the domain to retrieve the reconstructed values of the left- and right-going pieces of the flux at the cell boundaries, and the differences taken on either side of a cell face to determine the net flux at that face. 

To decompose the flux into purely left- and right-going pieces, we use the Lax-Friedrichs splitting,
\begin{equation}
\bm{f}_i^{\pm} = \frac{1}{2}\bigg(\bm{f}_i \pm \alpha \bm{q}_i \bigg)
\end{equation} 
where $\alpha$ is the maximum wave-speed across the reconstruction ($\alpha = c = 1$, in MHD), and $\bm{f}_i$ is the flux vector at the centre of cell $i$ due to the state given by $\bm{q}_i$, with $\bm{f}_i^{+} \ge 0$ and $\bm{f}_i^{-} \le 0$.

The values of the flux at the cell faces is generated using a reconstruction which prevents (limits) numerical oscillations at shocks. There are a number of options for this, including targeted, essentially non-oscillatory reconstructions \citep{Fu2017}, but we find that the third order WENO3 scheme of \cite{Shu1997} works well.

\section{METHOD}\label{sec:METHOD}

In this section we outline some of the design features of the relativistic and GPU-accelerated, multi-physics code METHOD.

METHOD is an object-oriented, C++ application for the evolution and testing of different MHD models. It is designed to allow maximum ease with which to test different features of plasma modelling using MHD, including flux reconstructions, time integrators, physics models and initial conditions, and to assess the feasibility of GPU-implementations to ease computational demands. Extending METHOD to allow execution on graphics cards is achieved using NVIDIA's CUDA, version 9.0, see section \ref{sec:CUDA}.

To ensure that tolerances of $10^{-13}$ are achieved for the primitive recovery, we have had to ensure that all data is stored in METHOD as a double precision, floating point number. This is in contrast to other GPU-capable MHD solvers such as \cite{Zink2011}, in which parallel speed-ups can be improved by relying only on single precision accuracy. As a result, this will limit the performance improvements that are possible with the more complex, multi-fluid models, as double precision is required to maintain stable evolutions.

\subsection{Parallel implementation}

To investigate the possibility of performance improvements by utilising graphics cards, the most computationally intensive tasks have been ported to the GPU. This form of design implementation is sub-optimal in the long term, as simulation data generally resides on the CPU and must be copied as needed to the GPU. This presents a limitation in the possible speed-up as host-to-device and device-to-host memory transfers are notoriously slow \citep{Bauer2011, Nickolls2008}. It does, however, ease the process of adapting existing code to become GPU-capable. A more computationally efficient design would be to keep all data on the GPU throughout the simulation and only copy data to the host for outputting. Memory limitations can be handled, along with additional performance improvements, by utilising multiple graphics cards. Data transfer between devices can then be done using CUDA-aware MPI calls. Future iterations of METHOD will be designed in such a way.

In order to achieve good parallel speed-ups by porting functions to the GPU, one needs to identify the bottlenecks in the simulation. For the two resistive models presented here, this is the root-finding procedure required by the IMEX schemes, and specifically the conservative-to-primitive variable transformation required in each iteration. We find that, depending upon the stiffness of the problem, an increase in execution time in excess of $10\times$ is not uncommon for resistive models using IMEX schemes when compared to ideal RMHD with explicit time-integrators.

To maximise the chances of performance improvements, we therefore port the time integrator, and any functions it implements, to the GPU. This includes the implicit root-find for the IMEX integrator, flux vector calculations, source vector calculations, and, crucially, the primitive recovery.

\subsection{CUDA overview}\label{sec:CUDA}

The power of general purpose GPU programming (GPGPU) lies in the number of floating point operations per second (FLOPS) permitted by modern day GPUs \citep{Nickolls2008}. Graphics cards typically have a few thousand cores per device---where a core is a single processor, capable of executing one instruction at a time. The immense number of cores available on a GPU means that they lend themselves very well to scientific programming, in which there are often a large number operations to be performed independently of each other. 

Cores are batched together on the GPU (or device) into groups of 32, called streaming multiprocessors (SMs). Batches of 32 threads, called warps, are executed concurrently in which the same instruction is applied by the threads in the warp on different data. Numbers of warps are further abstracted into blocks, and a number of blocks are dispatched on the device at a time. The distinction of threads into blocks, and blocks into grids eases thread recruitment and management of the data the threads are operating on. The optimum configuration of threads and blocks is discussed in section \ref{sec:optimisations}.

While it is easy to run a program on a graphics card, running the program efficiently and achieving good parallel speed-ups can be difficult. Problems that are suited to GPU-execution tend to require a large number of compute operations on a small data set, in addition to each operation being largely independent of each other. Whilst there may be some interdependence between threads (e.g.\ the flux reconstruction), fluid and MHD simulations are well suited to parallel programming architectures on account of the large amount of work required to evolve a single cell in the domain. 

\subsection{Host and device memory}

When programming for the CPU, developers rarely need to have an in-depth understanding of the underlying memory management. Modern day computers are extremely efficient at optimising access to data in host memory (RAM), and it is often sufficient to allocate memory and allow the CPU to automatically use caching to speed-up performance. The same cannot be said when considering GPGPU programming, however. 

NVIDIA graphics cards have multiple different kinds of memory spaces that the programmer must be familiar with in order to achieve efficient execution. Good performance improvements are achieved only when the correct balance of the various memory resources are used, and getting this step wrong can significantly hamper possible improvements. We will briefly explain some of the strategies applied in METHOD to optimise memory management. 

\subsubsection{Memory coalescence} \label{sec:memory coalescence}

When copying data between the CPU (or host) and device, one must ensure that the data is aligned in a suitable way. Transferring data to device memory essentially boils down to copying a single, one-dimensional array, and in order to optimise this process the developer must be familiar with how data is stored in memory and how the data is accessed for a computation. 

In METHOD, data is aligned in the following way: the conserved vector is an array with dimensions $N_{\text{cons}}\times N_x \times N_y  \times N_z$, which we can access using \lstinline|cons[var, i, j, k]| where \lstinline|var, i, j, k| access the \lstinline|var|$^{\text{th}}$ conserved variable in the \lstinline|i|$^{\text{th}}$, \lstinline|j|$^{\text{th}}$ and \lstinline|k|$^{\text{th}}$ cell in the $x$, $y$, $z$ directions, respectively. Due to the way in which memory is allocated in C++, all data is aligned first in the $z$-direction, followed by $y$, $x$ and then the variable, $var$---this means that all data in the $z$-direction is contiguous in memory. 

As the flux reconstruction requires interpolating data between neighbouring cells, this is easily performed in the $z$-direction as data is already contiguous in memory. Reconstructing in the $x$- and $y$-directions, however, requires us to rearrange the original array such that data is contiguous in the $x$- and $y$-directions, respectively. This rearranging of data is also necessary when copying data to the device for the IMEX integrator, as in this case, data must be contiguous in the conserved variables, $var$. 

The data is reorganised on the CPU so that it can be copied in a contiguous array to the device for operation. To reduce the performance hit this reorganising has, we use OpenMP \citep{OMP2015} to parallelise this process. Another possible option to reduce this overhead is to copy all data to the device in its original order and allow threads to copy the relevant data from global memory to shared memory into the correct order. Initial efforts to achieve this slowed simulations by many factors, and instead we opted to stick with the CUDA/OpenMP hybrid method. In section \ref{sec:validation and performance} we discuss the resulting performance impact. 

\subsubsection{Memory optimisation}\label{sec:optimisations}

There are three kinds of memory that are utilised in METHOD---global, shared and register---each with different sizes and latencies. Memory latency is the amount of time it takes from a thread requesting some data, to the data being available for computation. The different memory spaces have different latencies due to the distance from the where the threads operate, the SM, to the location of the data on the device.

Global memory is the largest block of memory, and likewise the slowest to access. When data is copied to the device from the host, it is copied into global memory where every thread can access it. Shared memory is smaller in size, typically $\approx 96$KB per SM, but around $100\times$ faster to access. Threads in the same block have access to the same shared memory, but cannot see data lying in another block's shared memory. This restriction means that when threads require data from their neighbours, one must overlap data between blocks. Finally, register memory is thread specific, and around $100\times$ quicker than shared memory.

Handling memory management well in GPGPU programming is the difference between order of magnitude speed-ups or slow-downs, and often a large amount of trial and error testing is required until an optimal balance between shared and global memory is found. The same can be said in terms of the configuration of threads and blocks---a large number of threads in a block will limit the amount of shared memory per block that is available, but a small number of threads may reduce the opportunity to hide memory latency behind useful computation. The optimal configuration will always be problem specific.

For the resistive RMHD models presented here, the best strategy we found is to maximise usage of shared memory for only 32 threads per block. This shared memory is specifically allocated for use in the primitive recovery, as these variables are accessed a very large number of times throughout the course of the IMEX scheme, and the benefits of having quicker access to these variables outweighs the benefits of having them lie in global memory with the possibility of hiding memory latency within blocks. In order to hide latency between blocks, we therefore choose the maximum number of blocks possible.

The most notable difference between implementing ideal and resistive MHD models on the GPU is the necessity of an implicit integrator to maintain stability in stiff regions. The process of implementing an implicit step requires some $N$-dimensional root-finding procedure, in which $N$ is the size of the conserved vector. The package used for this process in METHOD is adapted from CMINPACK\footnote{Original source: https://github.com/devernay/cminpack}, a C version of the widely used Fortran package, MINPACK. CMINPACK requires the use of a work-array of size no less than $N(3N+13)$ double precision, floating point numbers. The work-array provides the CMINPACK routines with resources to determine the numerical Jacobian of the system, error norms, etcetera, and so requires data that will be accessed multiple times throughout a single step. Ideally, this would lie in shared memory to minimise latency, but due to the size of both resistive systems and the limitations imposed by the size of the shared memory, the work-array must be allocated in global memory. This issue is specific to systems that require implicit integrators, and does not occur for ideal RMHD models.

\section{Validation and performance}\label{sec:validation and performance}
In this section we begin by validating the solutions generated by METHOD for some established MHD and fluid initial data. Following this, we present the performance improvements achieved by execution on different generations of NVIDIA graphics cards, and show that parallel speed-ups of well over an order of magnitude are possible. 

The range of initial sets ups available in METHOD extend beyond what is presented here---we only show some results, but note that the solutions agree with the literature in every case for all of the models implemented.

\subsection{Brio-Wu shock tube}
A common set-up that demonstrates shock-handling capabilities is the one-dimensional Brio-Wu shock tube test \citep{Brio1988}. The domain is divided into two states, separated by a discontinuity, and at $t=0$ the partition is removed, the two states begin to interact and a series of waves propagate through the system. This problem reduces to a single Riemann problem, and so there exists an exact solution with which we can compare our solutions. The exact solution in the ideal RMHD limit, $\sigma \to \infty$, is generated using the code of \citet{Giacomazzo2005}.

\subsubsection*{Single fluid}
To verify the behaviour in the single fluid, resistive regime we use the set up from \cite{Palenzuela2009}, and vary the magnitude of the conductivity. The initial state is given by Equations (\ref{eq: brio wu single initial}) and we use $\Gamma = 2$ for consistency with \cite{Palenzuela2009} and \cite{Dionysopoulou2013}. 

\begin{equation} \label{eq: brio wu single initial}
(\rho, p, B_y) = 
\begin{cases}
(1, 1, 0.5) &\text{for} \ 0 \le x < L/2, \\
(0.125, 0.1, -0.5) &\text{for} \ L/2 \le x \le L.
\end{cases}
\end{equation}

\begin{figure}[!h]
	\gridline{\fig{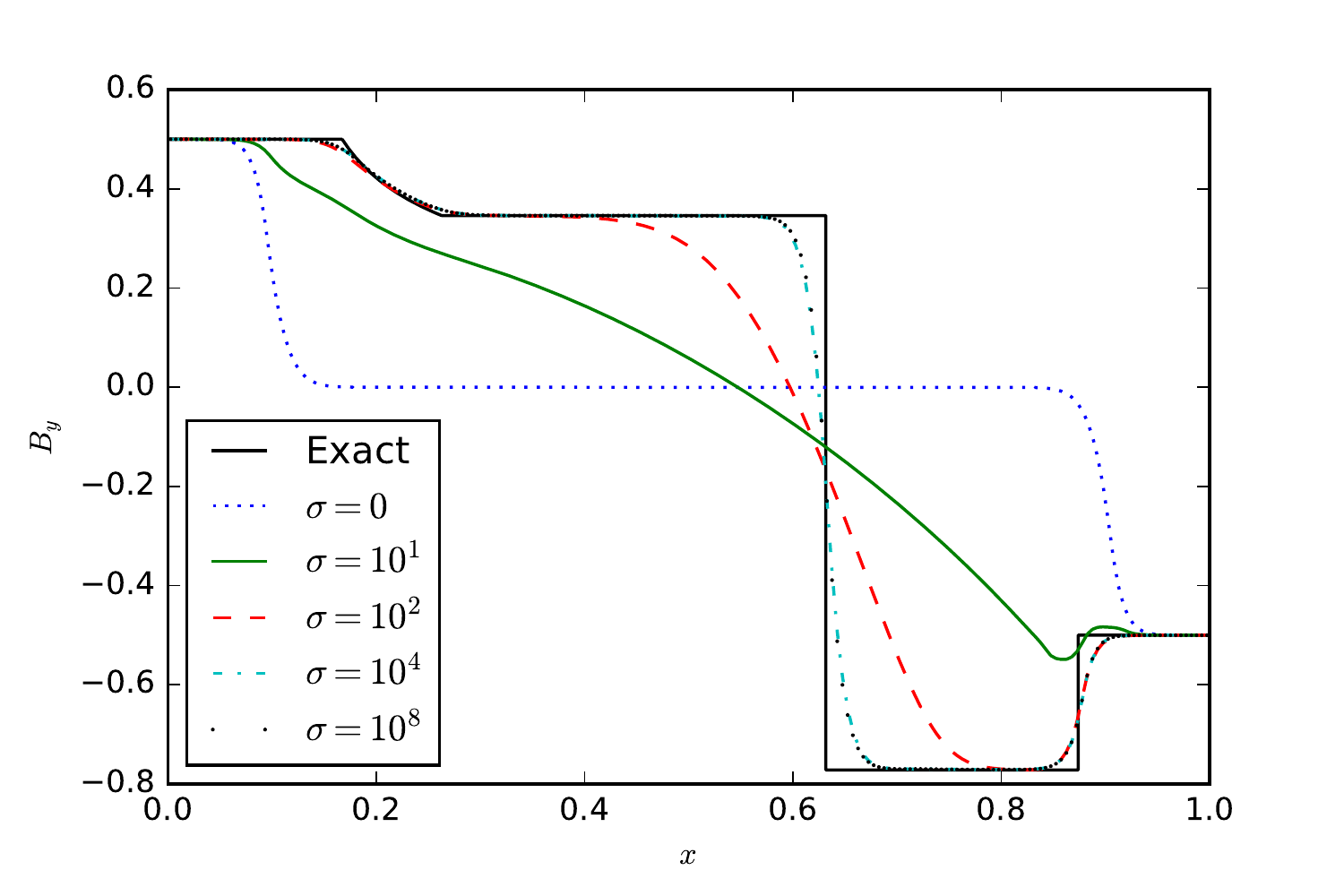}{0.6\textwidth}{}}
	\caption{$y$-component of the magnetic field for the single-fluid, resistive RMHD model using the Brio-Wu initial data---as $\sigma$ increases, the system tends towards the ideal RMHD solution. The system is evolved until $T=0.4$ for $N_x = 200$ cells.}
	\label{fig:brio wu resistive}
\end{figure}

\subsubsection*{Two-Fluid}
For the two-fluid RMHD model we use the initial data from \cite{Amano2016}, given by Equations (\ref{eq: brio wu twofluid initial}). At high enough resolutions, we can see the onset of the two-fluid effect as expected.

\begin{equation} \label{eq: brio wu twofluid initial}
(\rho, p, B_x, B_y) = 
\begin{cases}
(1, 1, 0.5, 1) &\text{for} \ 0 \le x < L/2, \\
(0.125, 0.1, 0.5, -1) &\text{for} \ L/2 \le x \le L.
\end{cases}
\end{equation}

Results for both resistive RMHD models are in excellent agreement with the literature, and show that METHOD can avoid numerical oscillations and instabilities for discontinuous data.

\subsection{Self-similar current sheet}

A useful test of the resistive nature of MHD models was given first in \cite{Komissarov2008}. The system is set in hydrodynamic equilibrium with $(p, \bm{v}, \rho) = (50.0, \bm{0}, 1.0)$. If a magnetic field of $(B_x, B_y, B_z) = (0, B(x, t), 0)$ is applied, the resulting dynamics of the field is given by the diffusion equation,
\begin{equation}
\partial_t B - \eta \partial_x^2B = 0,
\end{equation}
a solution of which is given by the self-similar error function:
\begin{equation}
B_y(x, t) = B_0 \ \text{erf} \bigg(\frac{\sqrt{\sigma x^2}}{2t}\bigg).
\end{equation}

\begin{figure}[!h]
	\centering
	\hspace*{-0.75cm}
	\label{fig:current sheet two fluid}
	\gridline{\fig{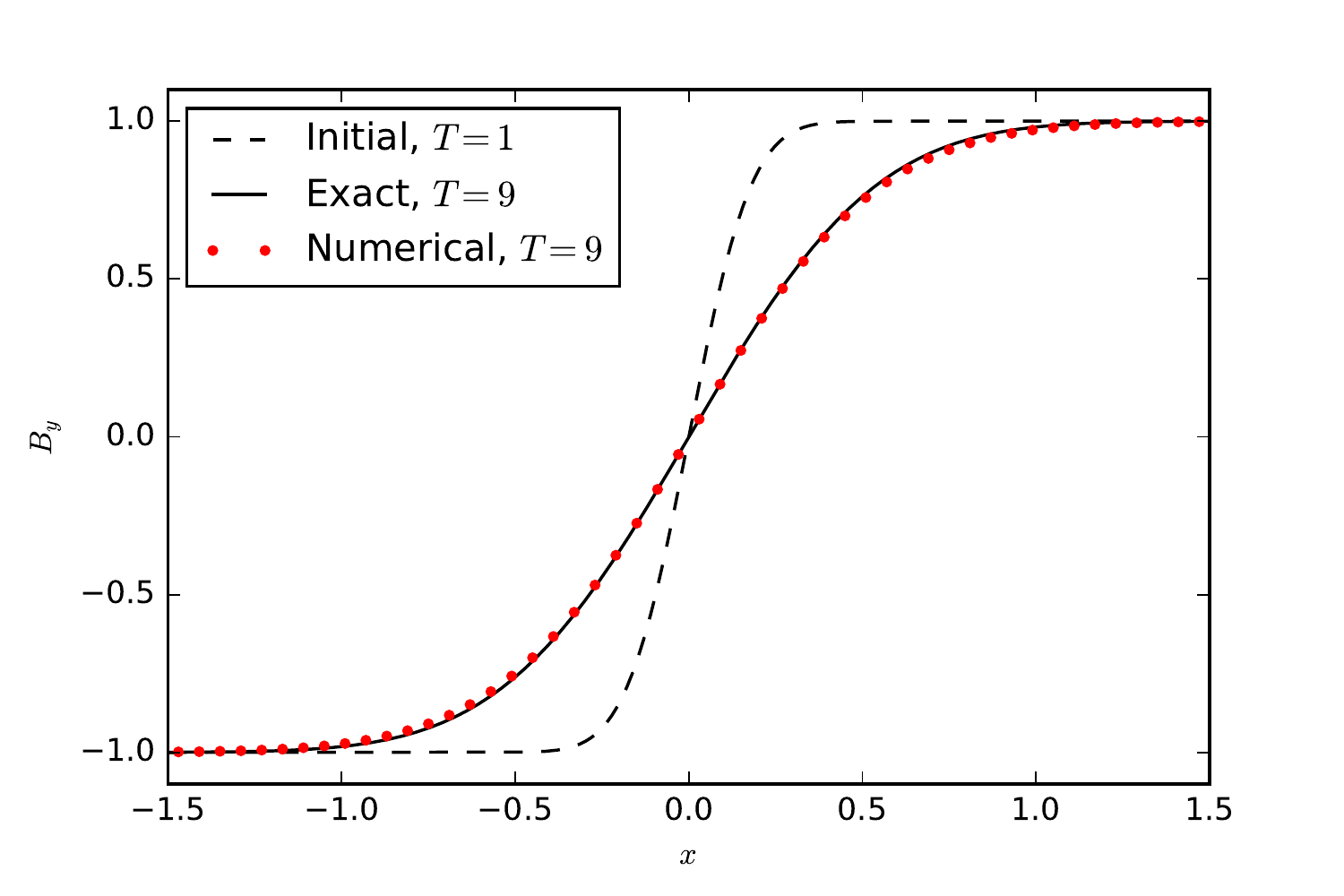}{0.6\textwidth}{}}
	\caption{Comparison of the exact and numerical solutions for the self-similar current sheet for two-fluid model using $N=50$ cells and simulating a pair plasma, $\mu_p = 10^3 = -\mu_e$.}
\end{figure}

At $t=1.0$, we initialise the electromagnetic fields with $(\bm{E}, B_0) = (\bm{0}, 1.0)$ so that the magnetic pressure is much less than the hydrodynamic pressure, and use a moderate conductivity $\sigma = 100$. The $B_y$ solution at $T=9.0$ is shown in figure \ref{fig:current sheet two fluid} for the two-fluid model for $N=50$ cells---even at such low resolutions the exact and numerical solutions are indistinguishable.

\subsection{Field Loop Advection}
Convergence tests are useful tools to assess code behaviour. We expect that as the resolution of the simulation is increased, the numerical solution should converge to the true solution as a function of the resolved scale. To show the numerical convergence of METHOD, we present the field loop advection test, adapted from \cite{Gardiner2008}. In this set-up, the fluid is set in hydrodynamic equilibrium and a weak magnetic field loop is advected through the system.  Comparing the solution after one crossing time to the initial data for a range of spatial resolutions gives an indicator of the order of convergence of the algorithm.

\begin{figure}[!h]
	\centering
	\plotone{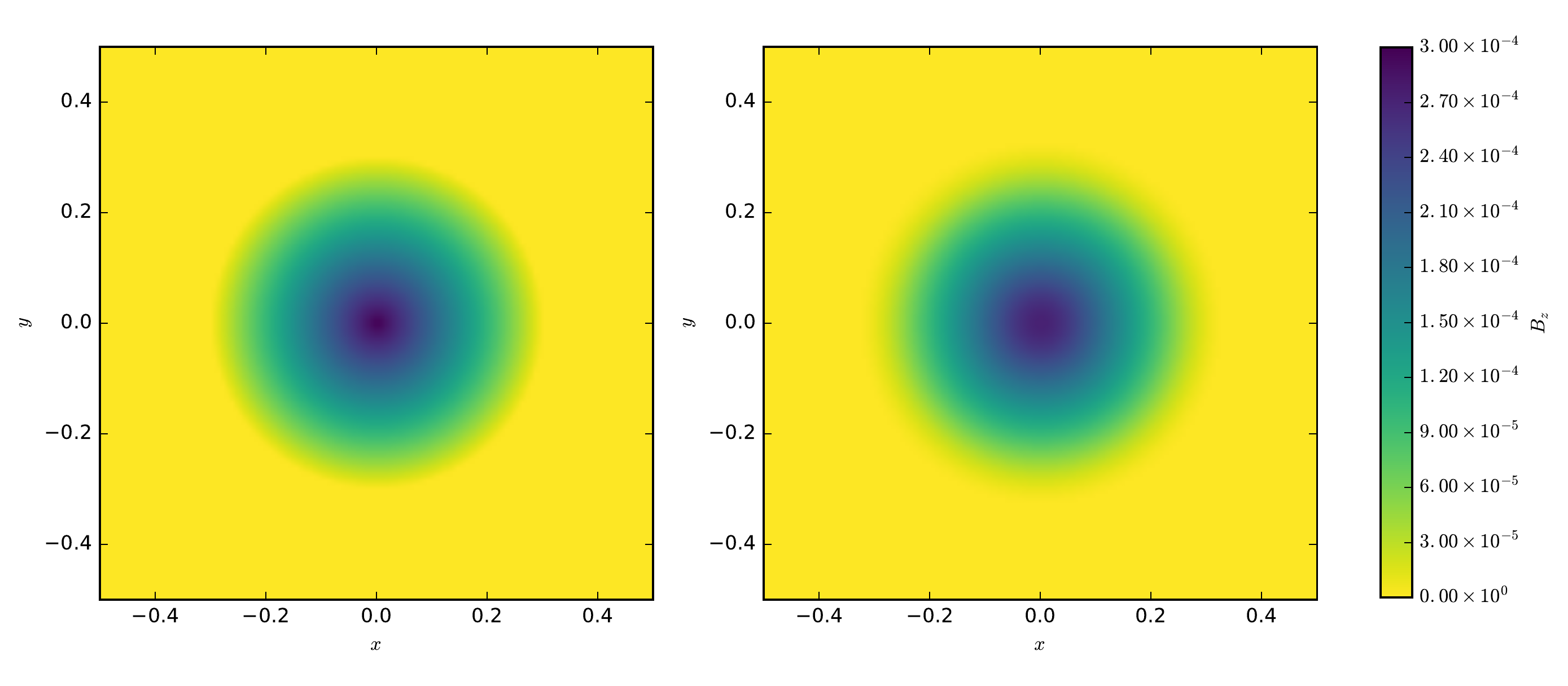}
	\caption{The initial state (left) and the final state after one crossing time for the z-direction magnetic field, $B_z$, using $128\times128$ grid points for the field loop advection test.}
	\label{fig:FieldLoopAdvectionFinal}
\end{figure}
\begin{figure}[!h]
	\centering
	\gridline{\fig{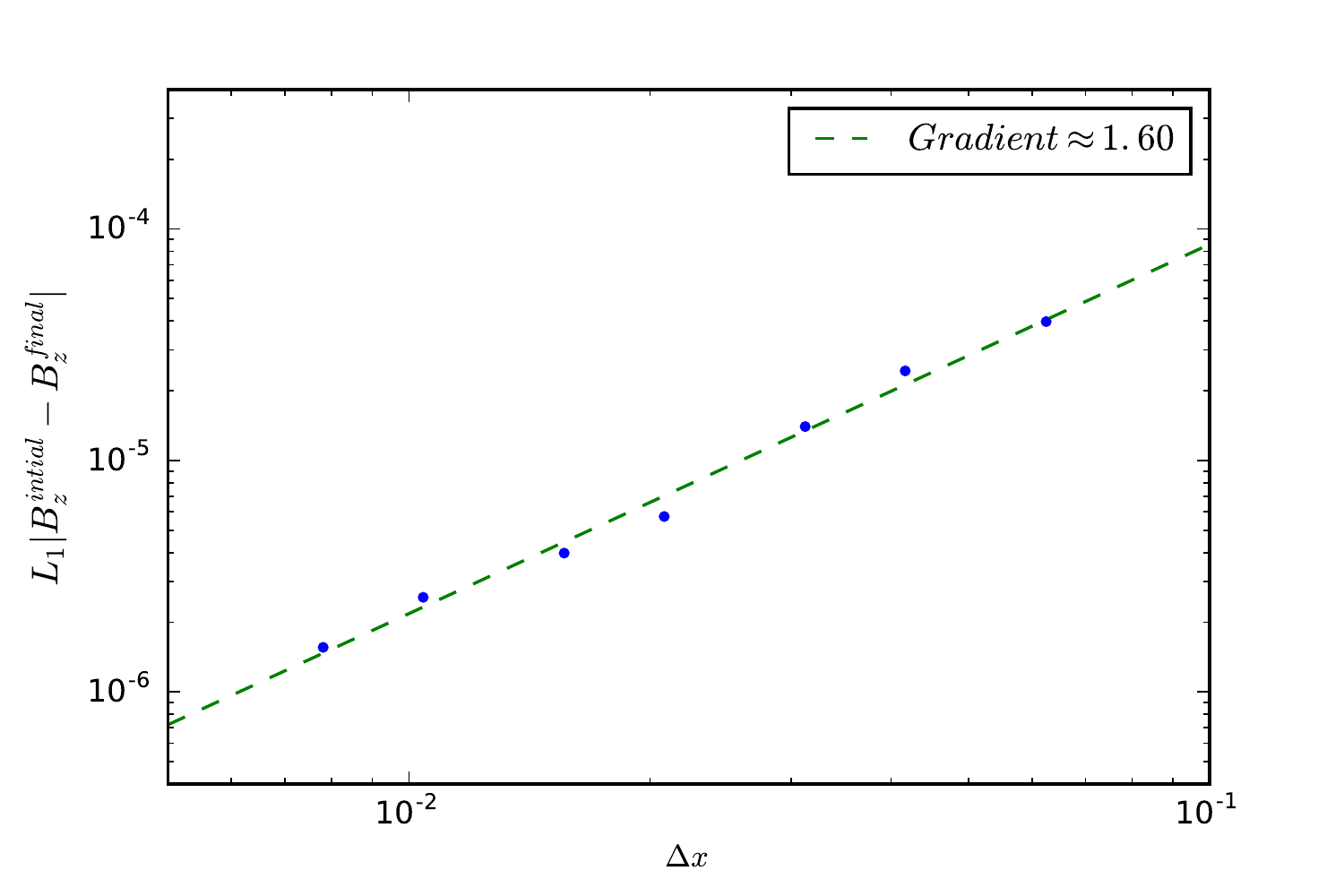}{0.6\textwidth}{}}		
	\caption{The convergence for a range of resolutions for the field loop advection test. Although the convergence is slightly less than expected, we can see that the numerical solution converges well with an increase in resolution.}
	\label{fig:FieldLoopAdvectionConvergence}
\end{figure}

The hydrodynamic system is initialised with $(\rho, v_x, v_y, v_z, p) = (3, 0.1, 0.1, 0.0, 1)$ in a two-dimensional domain with boundaries at $(-0.5, 0.5)\times(-0.5, 0.5)$, and composed of $N\times N$ grid points. The z-direction magnetic field is then set as
\begin{equation}
B_z = 
\begin{cases}
B_0 (R - r) &\text{ for } r \le R \\
0 &\text{ for } r > R
\end{cases}
\end{equation}
with $B_0 = 10^{-3}$. 

For this test, we used the SSP2(222) integrator and compared initial and final states after one crossing time, $T=10$, with periodic boundaries. The results are shown in figure \ref{fig:FieldLoopAdvectionFinal} for the single-fluid resistive model in the stiff limit, $\sigma=10^5$, and with $\Gamma=2.0$. The error is calculated with the $L_1$-norm for all cells in the system, and plotted as a function of the spatial resolution, where we note that $\Delta x = \Delta y$. The gradient is less than the expected value of 2 (which is limited by the order of the time integrator), figure \ref{fig:FieldLoopAdvectionConvergence}, but still shows strong convergence with increasing resolution.

\subsection{Kelvin-Helmholtz instability}
A common instability that occurs in real-life applications of numerical fluid codes, and believed to be instrumental in forming strong magnetic fields in mergers \citep{Kiuchi2015, Zhang2009, Obergaulinger2010}, is one of differentially flowing fluids, the so-called Kelvin-Helmholtz instability (KHI). This two-dimensional problem concerns the growth of perturbations along a boundary layer between two fluids with relative velocities, which evolve to the recognisable vortices seen in some cloud formations. 

\begin{figure}[!h]
	\centering
	\gridline{\fig{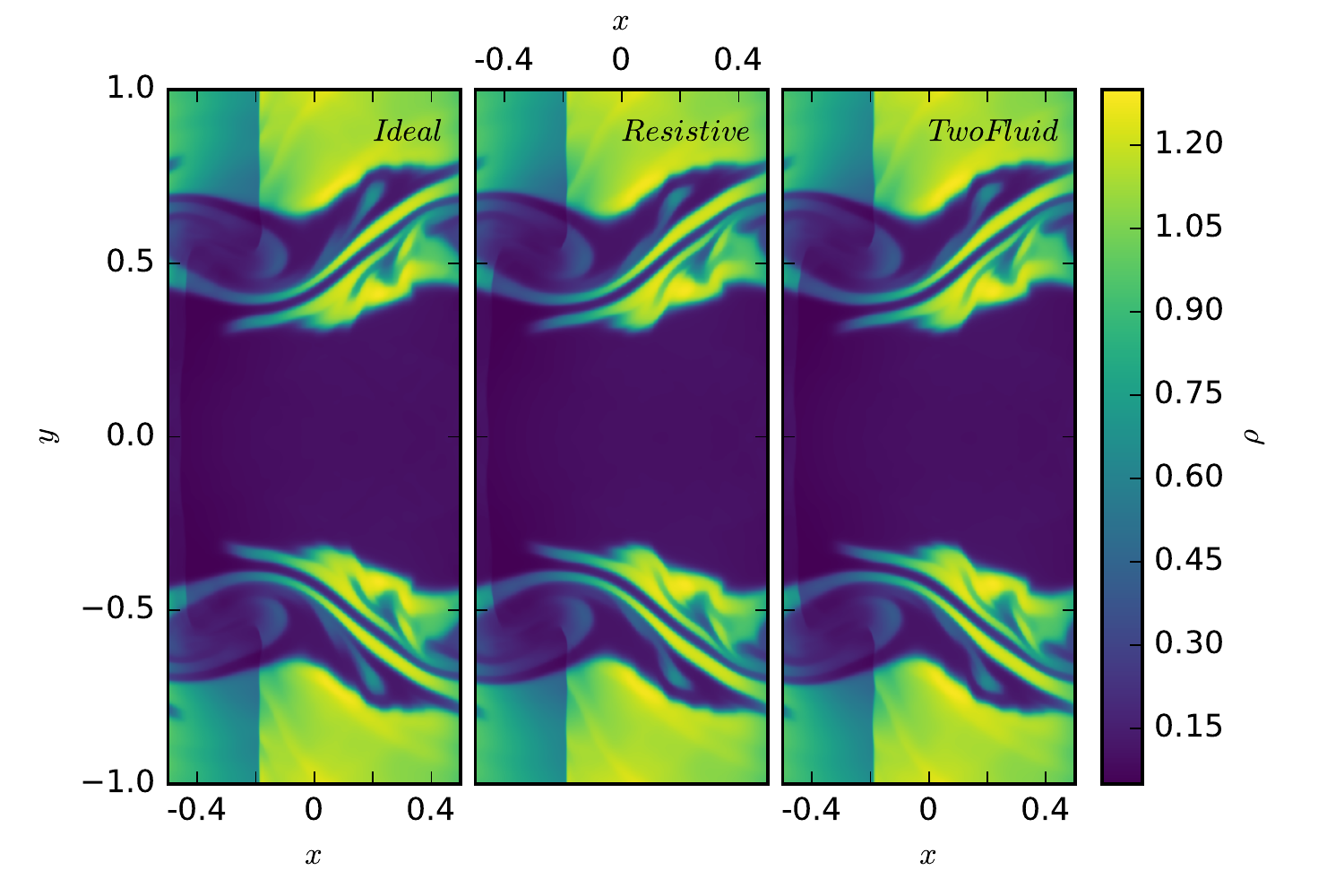}{0.49\textwidth}{(a)} 
					\fig{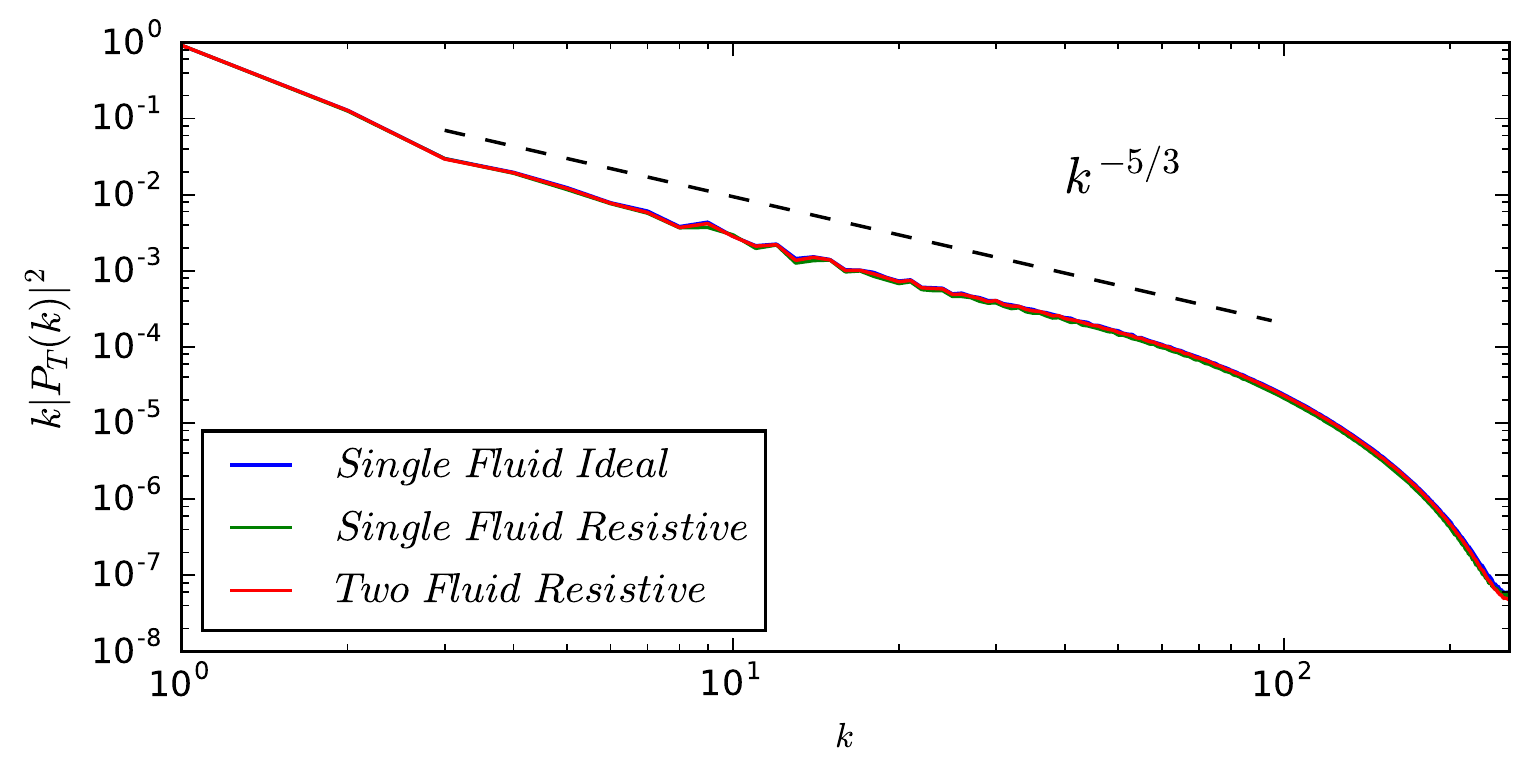}{0.49\textwidth}{(b)}}	
	\caption{Results for the Kelvin-Helmholtz instability simulation. Figure (a) shows the final state of the density for the ideal and resistive single- and two-fluid models at $T=6.0$, and Figure (b) the kinetic energy power spectrum for each model at $T=3.0$. All models follow Kolmogorov's 5/3 power law. The hydrodynamic behaviour for each model is virtually indistinguishable due to low magnitude of the magnetic fields and the relatively coarse resolution. All models are evolved with $512\times1024$ cells.}
	\label{fig:KHI}
\end{figure}

The initial data has been adapted from \cite{Mignone2009}. We initialise a two dimensional domain, where $x,y \in [-0.5, 0.5],[-1.0, 1.0]$, with the following properties:
\begin{equation}
v_x = 
\begin{cases}
&\ \ v_{shear} \tanh \big(\frac{y-0.5}{a} \big)	\ \ \ \text{if} \ \ y > 0.0 \\
&-v_{shear} \tanh \big( \frac{y+0.5}{a} \big) \ \ \ \text{if} \ \ y \le 0.0,
\end{cases}
\end{equation}
\begin{equation} \label{eq: KHI vy single fluid}
v_y = 
\begin{cases}
&\ \ A_0 v_{shear} \sin (2 \pi x) e^{\frac{-(y-0.5)^2}{l^2}}	\ \ \ \text{if} \ \ y > 0.0 \\
&-A_0 v_{shear} \sin (2 \pi x) e^{\frac{-(y+0.5)^2}{l^2}}	\ \ \ \text{if} \ \ y \le 0.0,
\end{cases}
\end{equation}
\begin{equation}
\rho = 
\begin{cases}
&\rho_0 + \rho_1 \tanh \big(\frac{y-0.5}{a} \big)	\ \ \ \text{if} \ \ y > 0.0 \\
&\rho_0 - \rho_1 \tanh \big( \frac{y+0.5}{a} \big) \ \ \ \text{if} \ \ y \le 0.0,
\end{cases}
\end{equation}
in which the shear velocity is $v_{shear}=0.5$ with a thickness of $a=0.01$, $(\rho_0, \rho_1)=(0.55, 0.45)$, and the amplitude perturbation in $y$-direction is $A=0.1$ over a length scale of $l=0.1$. The system is initially at a constant pressure of $p=1.0$ with $\Gamma = 4/3$ and a magnetic field perpendicular to the fluid flow of $B_z=0.1$, the conductivity is set at $\sigma = 10$, and a charge-mass ratio of $\mu_p = -\mu_e = 10^2$ is used for the two fluid model.

Figure \ref{fig:KHI} shows the density distribution at $T=6.0$ for each of the models (left). The magnetic field strength is low and so does not greatly impact on the behaviour of the fluid, resulting in near identical final states for all models. We also show the kinetic energy power spectrum, figure \ref{fig:KHI} (right), following the procedure in \cite{Beckwith2011} and note it follows Kolmogorov's 5/3 power law, as expected.

\subsection{Performance}

Our attention now turns to the performance of the CPU and GPU implementations of METHOD, and the potential benefits of GPU-capable, resistive MHD codes. For the comparison, we make use of the IRIDIS 5 HPC cluster at the University of Southampton---the GPU nodes possess a number of NVIDIA GTX1080 (Pascal architecture) (\cite{NVIDIA2016}) and V100 (Volta architecture) \citep{NVIDIA2017} graphics cards. We perform our analysis on a single GTX1080, and a single V100 card to assess performance differences between the two architectures, and compare these against a benchmark, CPU implementation on an INTEL XEON 6138 GOLD using a single core. 

\begin{figure}[!ht]
	\centering
	\hspace*{-0.75cm}
	\gridline{\fig{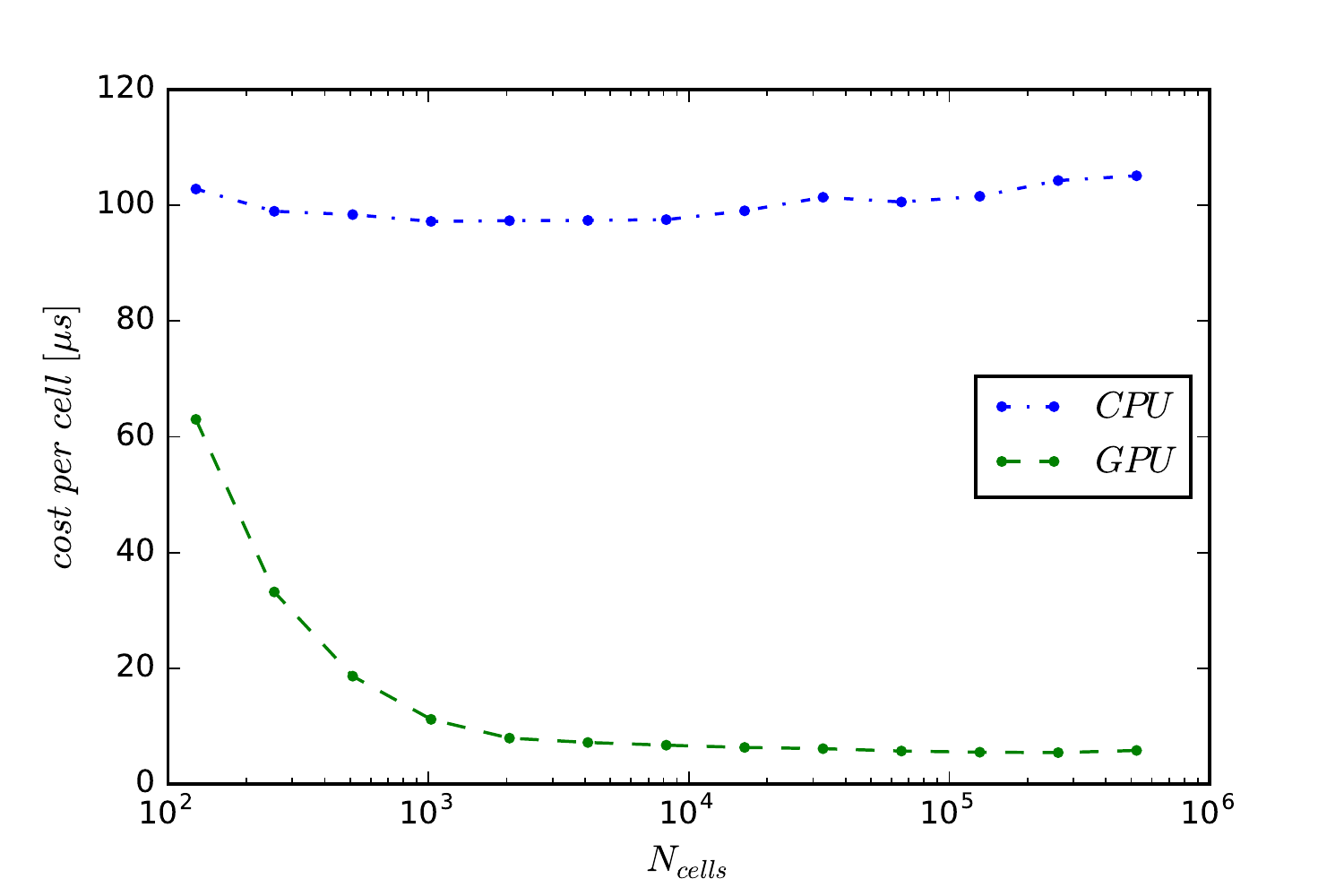}{0.5\textwidth}{}}
	\caption{A measure of the cost-per-cell (execution time in $\mu s$) on a V100 graphics card and the CPU for the Brio-Wu test using the SSP3 integrator. Execution on the CPU is independent of the domain size, whilst execution time on the device decreases until the GPU is saturated.}
	\label{fig:cost per cell}
\end{figure}

To compare the execution times of the serial and parallel implementations, we measure the wall clock time of the main evolution loop for ten iterations. This execution does not involve any time spent in setting up the domain or initial data, or writing data to the disk, and so is representative of the differences one would expect in the use of these methods for large-scale merger simulations where the majority of the execution time is spent on the evolution of the system (as opposed to initialisation or output of data).

We use two different initial data for the comparison---the Brio-Wu shock tube test and the more complex KHI instability, both described in the previous section. Furthermore, as we are only concerned with the GPU implementations of the two resistive models, we use the IMEX schemes mentioned in section \ref{sec:IMEX}---results from both the SSP2(222) and SSP3(322) integrators \citep{Pareschi2004} are presented. All simulations are run using double-precision floating-point accuracy as per the discussion in section \ref{sec:primitive recovery}. 

\begin{figure}[!ht]
	\centering
	\plottwo{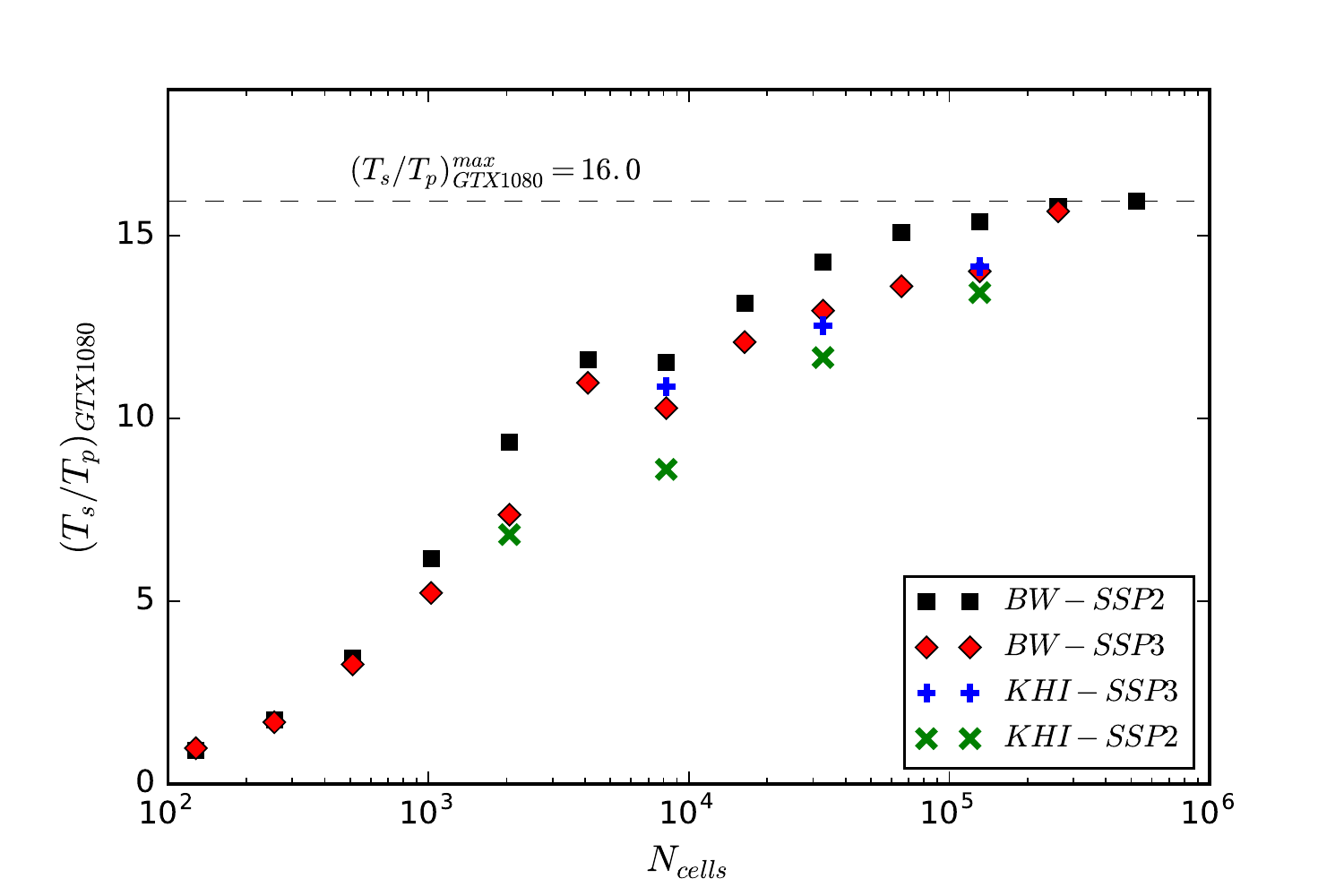}{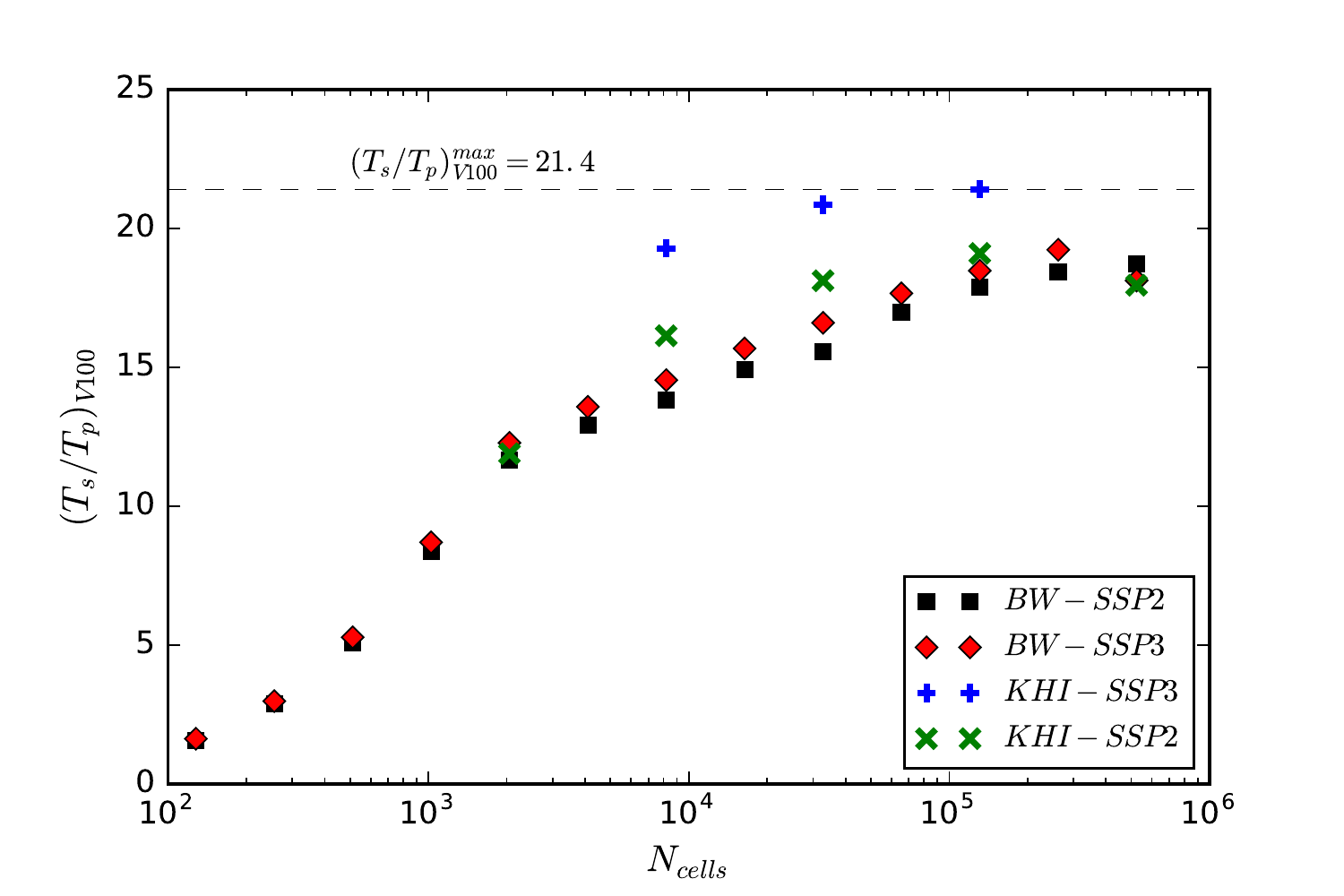}
	\caption{The measure of parallel speed-up, $T_{serial}/T_{parallel}$, for the GTX1080 (left) and V100 (right) graphics card. Squares and diamonds correspond to the SRRMHD single fluid Brio-Wu, while pluses and crosses to the two-fluid Kelvin-Helmholtz instability, respectively. The speed-up depends strongly upon the total number of cells, reaching a maximum of 21x before memory resources on the device are exhausted. The maximum parallel speed-up for each card is indicated by the dashed line.}
	\label{fig:parallel speedup}
\end{figure}

The serial execution time does not vary much between runs as differences depend only upon the current state of the CPU. The execution time for the parallel implementation, however, depends on a number of factors such as usage of shared memory and registers, number of host-device memory copies, and the configuration of threads and blocks. As such, presented here are the optimal configurations that minimise the execution time on the graphics card. We find that the optimal configurations is to minimise the number of threads in a block to 32, as discussed in section \ref{sec:optimisations}, to allow sufficient usage of shared memory for the primitive recovery. We then maximise the number of blocks per grid to hide latency, relax the condition that serial and parallel output must exactly match (whilst keeping physically valid results), compile device code with the \lstinline|-O3| flag and switch off CUDA's fused-multiply-add functionality. 

Figure \ref{fig:cost per cell} shows the average cost-per-cell for the Brio-Wu simulation when using the SSP3 integrator on the CPU and a V100. As expected, the cost-per-cell for the CPU implementation is independent of the size of the domain, all data lies in the same memory (RAM) and so accessing and computation of data is common among all cells. Instead, looking at the GPU implementation, we can see that as the number of cells in the simulation increases, the cost-per-cell reduces drastically. 

Smaller domains contain fewer cells, and thus less opportunity to hide memory latency behind useful computation. As a result, for small domains the problem is memory-bound---a significant amount of time is spent waiting for access to various locations in memory and as a result the threads lay idle. As we increase the size of the domain, the bottleneck then becomes the amount of computation required for each time step, and the problem is then compute-bound. We can see the transition of the problem from memory- to compute-bound in figure \ref{fig:cost per cell} as the cost-per-cell begins to plateau for the largest systems. At this point, the GPU is saturated, and improvements may only be made via smarter memory management, or additional GPUs.

To see how this impacts the absolute speed-up due to the GPU implementation, we plot the ratio of the serial execution time and parallel execution time (parallel speed-up) for ten time-steps as a function of the total number of grid points in the simulation, figure \ref{fig:parallel speedup}. The results for two generations of cards are shown, the GTX1080 and V100. It is clear by comparing the two cards that the newer generation graphics card, the V100 with the Volta architecture, provides improved speed-ups. It can also be seen that there is little difference between the results of the two initial data and the two IMEX schemes for the older (but still widely used) Pascal architecture. On the other hand, the new Volta architecture offers greater memory bandwidth (nearly $3\times$), and therefore a reduced memory latency---the result is to allow execution time to be dominated by computation. We see this as a much improved parallel speed-up, over $21\times$, specifically for the compute intensive algorithms of the two-fluid RMHD model with the SSP3-IMEX scheme.

The previous result suggests that higher order methods which require additional computation will achieve greater parallel speed-ups than lower order alternatives. The additional computation required by a IMEX-SSP3(433) scheme for example, which requires a total of four implicit stage and three flux reconstructions, would result in a greater proportion of the execution time being spent on the graphics card. As a result, the benefits of a GPU-implementation for simulations in which the higher order schemes are used would be more apparent---merger simulations typically employ fourth order methods, for example.

\begin{figure}[!ht]
	\centering
	\hspace*{-0.75cm}
	\gridline{\fig{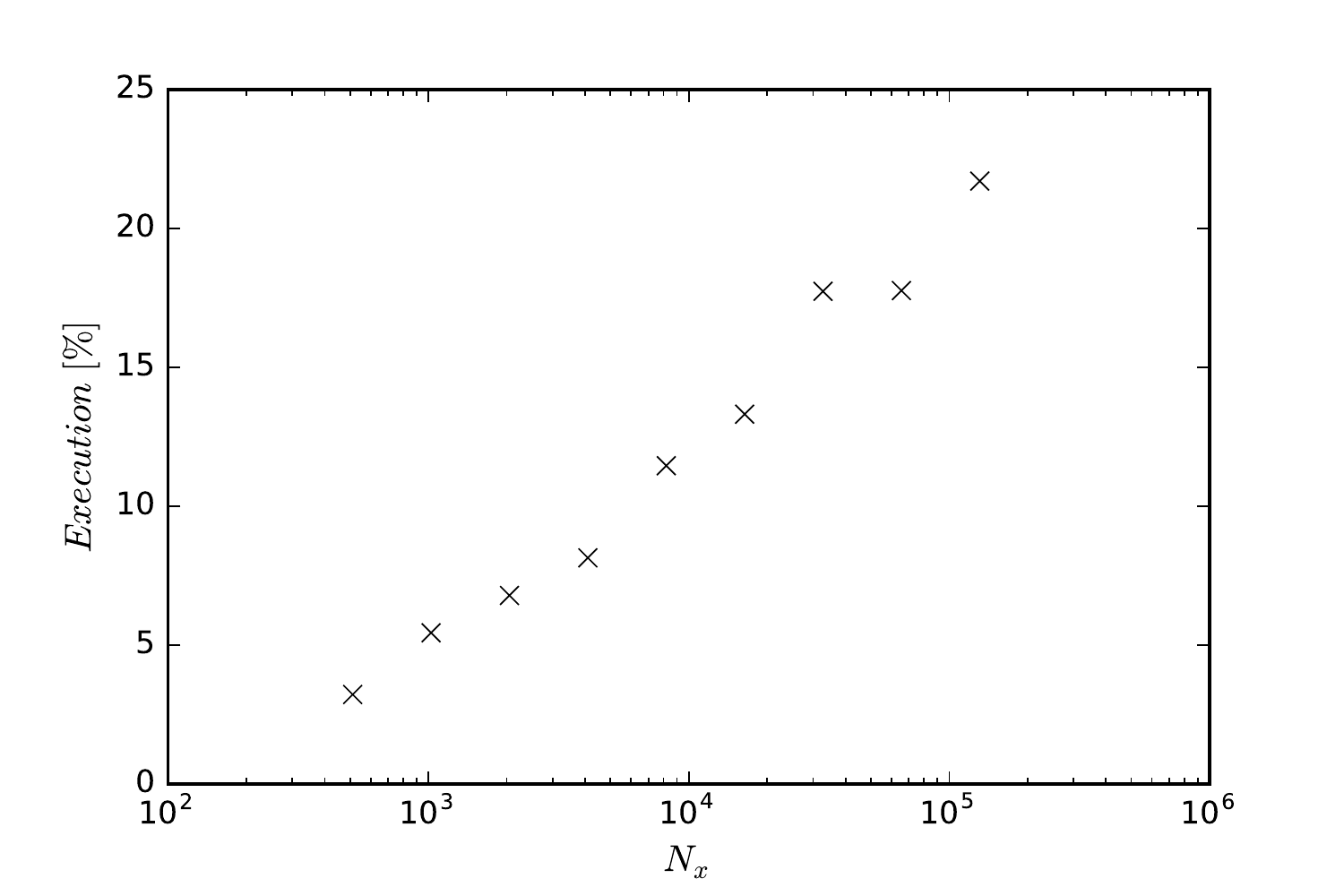}{0.5\textwidth}{}}
	\caption{Cost of rearranging data for IMEX scheme as a percentage of the total execution time. There is a strong dependence on the number of cells in the domain.}
	\label{fig:parallel profile}
\end{figure}

To gain insight into where the performance bottlenecks are in METHOD, we present a table with various profiling data, table \ref{table:profile}. Data was generated using the Brio-Wu initial data and a relatively large domain on $N_x=131072$ grid points, the SSP3 integrator using third order FVS reconstruction, and a conductivity of $\sigma=10^4$.  Even with execution on the device, the IMEX scheme makes up the largest proportion of the execution time. The next largest contributor is the rearrangement of data into contiguous arrays for the IMEX scheme, as discussed in section \ref{sec:memory coalescence}. The fraction of time taken to rearrange the data is heavily dependant upon the domain size. The expected scaling of the rearrangement on the CPU goes as the number of cells, $N_x$, but we have seen that the cost-per-cell of the main loop reduces with the number of cells. As a result, as the domain increases in size, more time is spent proportionally rearranging the data when compared to the total execution time (which is dominated by the execution on the device). We can see this visually in figure \ref{fig:parallel profile} as a growing fraction of the execution time being spent reordering data. Large scale simulations of neutron star mergers may evolve ten times as many grid points as we show here, and so a significant fraction of the execution time would be performing this rearrangement---in this case, a more efficient scheme for performing this would be highly beneficial.

Time spent rearranging data for the flux reconstruction, however, is significantly less than that for the IMEX scheme. As only the conserved and flux vectors need rearranging, this results in a total of 3 arrays being reordered in the FVS method. This is compared to a total of 28 arrays required for the SSP3 scheme, including the conserved, primitive, source and flux vectors for each stage. As a result, there are fewer benefits to be had from optimising the flux reconstruction. 

\setcounter{table}{2}
\begin{table}[h!]
	\renewcommand{\thetable}{\arabic{table}}
	\centering
	\caption{Performance profile for the Brio-Wu initial data using the SSP3 integrator and $N_x=2^{17}$ cells. Results represent the percentage of the total execution time of the main loop. The rest of the execution time is composed of the correction stage, Equation (\ref{eq:correction stage}), primitive recovery that is not within the implicit stages of the IMEX scheme, and data and memory management.}
	\label{table:profile}
	\begin{tabular}{ |c||c|c|c|c|c|}
	\hline
	\multicolumn{6}{|c|}{Performance Profile} \\
	\hline	
	\hline
	Functionality & IMEX rearrange & IMEX root-find & FVS rearrange & FVS reconstruction & boundary conditions\\
	\hline
	Proportion of execution [\%]   & 21.7 & 55.3 & 1.0&   2.2 & 0.2 \\
	\hline
\end{tabular}
\end{table}

\section{Discussion}\label{sec:discussion}

This work describes the development of a new, publicly available, GPU-accelerated, numerical, multi-physics RMHD code, METHOD. We have demonstrated the validity of its solutions, and investigated the available performance improvements made possible by porting compute intensive functions to the GPU. 

When evolving resistive single or multi-fluid RMHD models, one must use a time integrator that solves, at the least, the source terms implicitly---as a result, the majority of the execution time for these kinds of simulations is spent in these schemes. We have shown that the most commonly used class of integrators for this procedure, IMEX-RK schemes, can be accelerated by at least a factor of $21\times$ by execution on the device. Furthermore, we argue that there are potentially greater improvements to be made for higher order methods, as proportionally more time will be spent in the accelerated regions of the simulation. 

The greatest improvements of performance are possible when managing the resources of shared memory correctly. The recovery of the primitive variables, necessary for each iteration of the implicit stage in the IMEX scheme, itself requires a root-finding procedure. To minimise memory latency, the arrays required for the recovery are kept in shared memory---this severely limits the number of threads per block that we can recruit but saves multiple factors in terms of execution time.

The available speed-ups of METHOD are limited by its design---METHOD has been extended to execute the most computationally expensive functions (time integration) on the device, and inherent in this is a number of host-to-device and device-to-host memory transfers. In order to copy data from contiguous arrays we are required to perform a rearrangement of the data on the host which takes up a significant fraction of the overall execution. To reduce this overhead, and to improve the potential parallel speed-ups, we note that a more efficient design would be to develop the GPU-capabilities from the outset. In this way, all data will be in device memory and only transferred to the host for output. Organising the simulation data in this way may allow threads to load data into contiguous arrays in parallel on the device, which  for large domains may reduce the cost of the rearrangement significantly. 

Furthermore, even with the acceleration of the time-integrator by execution the device, the majority of time is still spent in the IMEX root-find. This means that optimisations in this region will likely produce further improvements. This could include reducing the size of the work-array required by the root-finding procedure such that it can fit in shared memory, although greater improvements seem likely by keeping the primitive recovery variables in shared memory at the expense of the IMEX work-array.

The size of the conserved vector has little effect on the possible speed up. The single-fluid RMHD model, which has the smallest conserved vector presented, requires a work array for the time integration that is too large to fit in shared memory. As a result, the speed-up comes primarily from the primitive recovery variables lying in shared memory. The same is true for models with larger conserved vectors, and so more complex, four-fluid models should expect comparable accelerations---the required memory for the primitive recovery grows linearly with the conserved vector size, in contrast to the quadratic dependence for the time integrator. As a result, the methods employed here should transfer over to the more computationally demanding, general-relativistic multi-fluid models proposed in \cite{Andersson2016}.

In the future, we aim to combine the results of what has been presented here with adaptive mesh refinement (AMR) techniques to utilise multiple GPUs in an MPI-OpenMP-CUDA hybrid code. Resistive multi-fluid models present us with a huge step in computation required to evolve a system. The increased size of the conserved vector and the necessity of implicit-explicit integrators for stable evolution mean that current implementations are not suitable for large scale, merger simulations. With a hybrid code of this type, one can utilise the immense amount of compute power provided by GPU clusters, allowing the evolution of more physically accurate systems than have previously been possible.

The first release of METHOD used for generating these results is available through Zenodo  \citep{Wright2018}, and the latest iteration is publicly available on GitHub\footnote{https://github.com/AlexJamesWright/METHOD}.

\acknowledgments

	AJW thanks Anthony Morse, Philip Blakely and Timothy Lanfear for insightful discussions about CUDA implementation and optimisation strategies. 
	The authors acknowledge the use of the IRIDIS High Performance Computing Facility, and associated support services at the University of Southampton, in the completion of this work. We also gratefully acknowledge financial support from the EPSRC Centre for Doctoral Training in Next Generation Computational Modelling grant EP/L015382/1.
	All plotting used the open-source Python package \textit{matplotlib} \citep{Hunter2007}.
	
\appendix

\section{Two-fluid model} \label{sec:two fluid features}
The flux vector and source term for the two-fluid, resistive RMHD model are as follows:
\begin{align} \label{eq:twoFluid flux}
\bm{f}^j(\bm{q}) = 
\begin{pmatrix}
\rho_s W_s v_s^j \\ S_{s,i}^j \\ S^j - D_s v_s^j \\
\overline{D}_s v_s^j \\ \overline{S}_s^j v_{s,i} + \delta^j_i p_s \mu_s \\ (\overline{\tau}_s + \mu_s p_s) v^j_s \\
\epsilon^{ijk} E_k + \delta_i^j \phi \\ -\epsilon^{ijk} B_k + \delta_i^j \psi \\ E^j \\ B^j
\end{pmatrix},
\end{align}

\begin{align} \label{eq:twoFluid source}
\bm{s}(\bm{q}) = 
\begin{pmatrix}
0 \\ 0 \\ 0 \\ 0 \\ \omega_p^2 \big[ W E_i + \epsilon_{ijk} u^j B^k - \eta (J_i - \varrho u_i) \big] \\ \omega_p^2 \big[ u_i E^i - \eta (\varrho - \varrho_0 W) \big] \\ 0 \\ - J^i \\ \varrho - \kappa \psi \\ - \kappa \phi
\end{pmatrix},
\end{align}
with the relations
\begin{align} \label{eq:twoFluid vars}
\begin{pmatrix}
S_{s,i}^j \\ \varrho  \\ J_i \\ W \\ u^i \\ \varrho_0 \\ \omega_p^2
\end{pmatrix} = 
\begin{pmatrix}
\rho_s h_s W_s^2 v_{s,i} v^j + \delta_i^j p_s - [E_i E^j + B_i B^j] + \delta_i^j [E^2 + B^2] / 2 \\ \mu_s \rho_s W_s \\ \mu_s \rho_s W_s v_{s,i} \\ \mu_s^2 \rho_s W_s / \omega_p^2\\ \mu_s^2 \rho_s W_s v_s^i / \omega_p^2 \\ W \varrho - J_i u^i \\ \mu^2_s \rho_s
\end{pmatrix}.
\end{align}

\end{document}